\documentclass[preprintnumbers,amsmath,amssymb,floatfix,superscriptaddress,nofootinbib]{revtex4}
\pdfoutput=1
\usepackage{amsmath,hyperref,amssymb}
\usepackage{graphicx}% Include figure files
\usepackage{dcolumn}% Align table columns on decimal point
\usepackage{bm}% bold math
\usepackage{verbatim}
\usepackage{tabularx}
\usepackage{slashed}
\usepackage{cancel}
\usepackage[boxsize=2em,aligntableaux=center]{ytableau}
\usepackage{float}

\def\be{\begin{equation}}
\def\ee{\end{equation}}
\def\bea{\begin{eqnarray}}
\def\eea{\end{eqnarray}}

\begin{document}

\title{Baryogenesis from Hawking Radiation}

%Non-supersymmetric S dualities
% S-duality of nonsupersymmetric gauge theories
% Nonsupersymmetric S dualities in four dimensional gauge theories

\author{Anson Hook}
\affiliation{School of Natural Sciences,
 Institute for Advanced Study\\
Princeton, NJ 08540, USA}


\begin{abstract}

We show that in the presence of a chemical potential, black hole evaporation generates baryon number.  
If the inflaton or Ricci scalar is derivatively coupled to the B-L current, the expansion of the universe acts as a chemical potential and splits the energy levels of particles and their anti-particles.
The asymmetric Hawking radiation of primordial black holes can thus be used to generate a B-L asymmetry.
If dark matter is produced by the same mechanism, the coincidence between the mass density of visible and dark matter can be naturally explained.
%A natural source for primordial black holes during inflation comes from the Higgs field.  The de Sitter temperature fluctuations of the Higgs fields can push it over its instability barrier and cause regions of space to crunch into black holes.  

\end{abstract}

%If we want pacs, need to add "showpacs" to \documentclass[]
%\pacs{}% PACS, the Physics and Astronomy
\maketitle

\section{Introduction}

The universe is observed to contain more matter than antimatter.  The baryon to entropy ratio has been measured to be $n_B/s = 9 \times 10^{-11}$.  Sakharov\cite{Sakharov:1967dj} formulated three conditions for the generation of the baryon number asymmetry.  The three conditions are baryon number violation, C and CP violation, and finally a departure from thermal equilibrium.  These conditions have guided the field to consider baryogenesis by the CP and baryon number violating decays of frozen out fields\footnote{See e.g. Ref.~\cite{Cline:2006ts} for a review and references.}.

In this note, we propose using the evaporation of black holes to generate the baryon number asymmetry\footnote{The idea of using black holes for baryogenesis dates back to Hawking\cite{Hawking:1974rv,Carr:1976zz,Baumann:2007yr}, where Hawking radiation is used as a source for producing GUT scale particles.  The subsequent CP and B-L violating decay of these GUT scale particles generated the baryon number asymmetry.  
See \cite{cite me} for other work in generating asymmetries in the early universe.}.  A ``folk theorem" is that black holes break all global symmetries~\cite{Kallosh:1995hi}.  The reasoning is that if a global symmetry is thrown into a black hole, then the subsequent thermal decay of the black hole destroys the global charge.  In the context of global symmetries, black holes are objects which break a global symmetry but decay symmetrically.  In this manner, black holes are similar to particles such as right handed neutrinos, which break lepton number, but also decay symmetrically into the decay channels $W^+ l^-$ and $W^- l^+$\footnote{In making this statement, we have neglected the small CP violation present in the Standard Model.}.

A simple way in which a symmetric decay can be made asymmetric is to change the kinematics.  If there is a chemical potential, then particles and anti-particles have different energy levels.  As a simple example, consider the decay of a right handed neutrino in the presence of a chemical potential for lepton number.  Conservation of energy dictates that
\bea
m_{\nu_R} = E_{W^\pm} + E_{l^\mp} \pm \mu
\eea
for the two different decay channels.  We see that if the chemical potential is larger than the mass of the right handed neutrino, one of the two decays is completely forbidden by energy conservation!  The presence of a chemical potential has made a symmetric decay 100\% asymmetric.  As black holes are also objects that break a symmetry but decay symmetrically, a chemical potential should also make black holes decay asymmetrically.

Black holes destroy global symmetries because their thermal evaporation results in equal numbers of particles and anti-particles.  
In the presence of a chemical potential, thermal systems carry global charges.  Thus the thermal nature of a black hole, which is usually used to argue that black holes destroy global charges, can also be used to generate global charges!  The idea of using a chemical potential to make black holes decay asymmetrically is not new.  This is the exact mechanism by which charged black holes radiate away their charge.  The electric potential couples as $A_0 J^0$ and acts like a chemical potential.  Thus when charged black holes decay, they decay thermally with a chemical potential $\mu = q A_0$ evaluated at the horizon~\cite{Hawking:1974rv}.  

In this note, we consider a dynamically generated chemical potential.  Chemical potentials break CPT.  A dynamical chemical potential will be generated by exploiting the expansion of the universe, which breaks time translational invariance, and combining it with a CP violating coupling.  There are two CP violating interactions that will be used in this paper.  
The first CP violating interaction is the coupling used in gravitational baryogenesis\cite{Davoudiasl:2004gf}
\bea
\label{Eq: def2}
\mathcal{L} \supset \int dx^4 \sqrt{-g} \lambda \frac{\partial_\mu \mathcal{R}}{M_p^2} J^\mu_{B-L}
\eea
where $\mathcal{R}$ is the Ricci scalar.  Due to the expansion of the universe, $\mathcal{\dot R} \ne 0$ so there is a chemical potential for the $B-L$ current.   Eq.~\ref{Eq: def2} violates CP so that when $\mathcal{\dot R}$ is non-zero, it breaks CPT and favors the production of particles over anti-particles or visa versa depending on the sign.  The breaking of CPT allows for B-L to be generated while in thermal equilibrium, violating one of Sakharov's three conditions.

The second CP violating interaction is similar to spontaneous baryogenesis\cite{Cohen:1987vi}.  If the inflaton has an approximate shift symmetry, then the most relevant operator coupling the Standard Model to the inflaton is
\bea
\label{Eq: def}
\mathcal{L} \supset \int dx^4 \sqrt{-g} \frac{\partial_\mu \phi}{M_\star} J^\mu_{B-L}
\eea
The two couplings are related by $\partial_\mu R \sim (H^2 / M_p) \partial_\mu \phi$.  As Eq.~\ref{Eq: def} is enhanced by the ratio $M_p^2/H^2$ as compared to Eq.~\ref{Eq: def2}, it is the dominant effect during inflation.  After inflation, Eq.~\ref{Eq: def2} is the dominant contribution to the chemical potential.  Thus the asymmetric evaporation of black holes during inflation is dominated by Eq.~\ref{Eq: def} while after inflation it is dominated by Eq.~\ref{Eq: def2}.

When integrated by parts, Eq.~\ref{Eq: def2} and Eq.~\ref{Eq: def} show that if $J^\mu_\text{B-L}$ is conserved, then up to boundary terms, the chemical potential has no effect. 
If $B-L$ is conserved, no net charge is generated.  To generate B-L, Ref.~\cite{Cohen:1987vi} postulated the existence of a baryon number violating operator.  In thermal equilibrium, this baryon number violating operator would convert the higher energy anti-particles into the lower energy particles generating a net B-L number.  

In our approach we use the fact that quantum gravity does not respect non-gauged symmetries.  In the presence of black holes, the outside observer cannot see past the horizon.  Thus when integrating by parts, one picks up a boundary term due to the black holes.  Since black holes do not respect global symmetries, it can inject particles asymmetrically into the universe.   We will show that because a chemical potential favors particles over anti-particles, the thermal evaporation of black holes can generate a baryon number asymmetry.  Despite the fact that B-L is a good perturbative symmetry of the theory, quantum gravity can still generate a baryon number asymmetry.

We stress that we assume that B-L is a good symmetry of all perturbative physics, even higher dimensional operators.  Thus to all orders in perturbation theory, Eq.~\ref{Eq: def2} and Eq.~\ref{Eq: def} have no effect.  It is only non-perturbative objects, like black holes, that are effected by the non-zero couplings of these terms.  Black holes serve as a bath which injects B-L charge into the system to minimize the energy.  This mechanism works even in the context of the Standard Model (SM) where B-L is a good symmetry of the theory.

This approach towards generating an asymmetry is very familiar from physical experiments.  In laboratory experiments, when a system has a chemical potential applied to it, it draws a number density to it from the bath surrounding it.  The resulting system has a number density of particles in the interior and an equal number of anti-particles in the surrounding bath.  In this case, the black holes provide a thermal bath from which to draw particles from.  As the outside observer measures only the universe outside of the horizon, he observes a non-zero B-L number.

This mechanism naturally accounts for dark matter as well\footnote{See \cite{Jungman:1995df} and \cite{Zurek:2013wia} for a review of (asymmetric) dark matter.}.  In most models of dark matter, there is a symmetry under which the the dark matter is the lightest particle.  If this symmetry is a global symmetry, e.g. a $U(1)_D$, then the coupling
\bea
\label{Eq: DMchem}
\mathcal{L} \supset \int dx^4 \sqrt{-g} \frac{\partial_\mu \phi}{M'_\star} J^\mu_{D}
\eea
naturally generates dark matter with comparable abundance to the baryons.  If the dark matter has mass in the GeV range, it naturally explains why $\Omega_m$ is so similar to $\Omega_D$.

Sec.~\ref{Sec: Black holes} demonstrates that in the presence of a chemical potential, Hawking radiation generates B-L charge.  Sec.~\ref{Sec: after inflation} shows how the evaporation of primordial black holes after inflation can be used to obtain the observed baryon number asymmetry.  Sec.~\ref{Sec: baryogenesis} shows how the evaporation of primordial black holes during inflation can also be used to obtain the observed baryon number asymmetry.    Finally, Sec.~\ref{Sec: production} briefly mentions production mechanisms for primordial black holes.

\section{Generation of B-L from the evaporation of black holes}
\label{Sec: Black holes}

In this section, we consider the evaporation of black holes in the presence of a chemical potential($\mu$) at its horizon\footnote{The chemical potentials in Eq.~\ref{Eq: def2} and Eq.~\ref{Eq: def} will generically have a different value at the horizon than asymptotically far away.  For this paper, we'll make a simplifying assumption that they are constant across all space.}.  The simplest argument for the evaporation of black holes remaining thermal in the presence of a chemical potential comes from analytically continuing black holes to Euclidean space\cite{Gibbons:1976ue}.  Requiring the absence of deficit angles gives the correct temperature of the black hole.  In this approach, the temperature of the black hole is derived without any reference to the matter content or chemical potential.  As such, Hawking radiation should remain thermal in the presence of arbitrary matter content and chemical potentials.

Hawking radiation acting like there is a chemical potential is not a new phenomenon\cite{Hawking:1974sw,Robinson:2005pd,Iso:2006wa}.  Hawking originally showed that the Hawking radiation of charged black holes behaves as if there is a chemical potential of size $e \Phi = e Q_\text{BH}/r_\text{BH}$, where $\Phi = A_0$ is the electric potential evaluated at the horizon of the black hole.  The chemical potential for charged spinning black holes results in the well known phenomenon of superradiance\cite{Gibbons:1975kk}.  This chemical potential is how a charged black hole radiates away its charge.  

In the case of a charged black hole, one analytically continues the geometry into euclidean space and finds the Hawking temperature.  Armed with the Hawking temperature, we say that the charged black hole decays with a temperature $T_\text{Hawking}$.  Due to the electric charge of the black hole, there is a non-zero electric potential at the horizon.  Because the gauge field couples as $A_\mu J^\mu$, this coupling acts as a chemical potential.  Thus the charged black hole decays with a temperature $T_\text{Hawking}$ with a chemical potential $\mu = e Q_\text{BH}/r_\text{BH}$.  This is the result first found by Hawking~\cite{Hawking:1974sw}.  The case of a black hole with an explicit chemical potential follows from the exact same arguments.  The only difference is that an explicit chemical potential, unlike an electric field, does not back react on the geometry and change the temperature.

\subsection{The charged black hole}

As a warm-up for an explicit chemical potential, we briefly cover how a black hole with a small charge loses its energy and charge.  A charged black hole has the metric
\bea
ds^2 = - (1 - \frac{2 M}{r M_p^2} + \frac{Q^2}{r^2 M_p^2}) dt^2 + (1 - \frac{2 M}{r M_p^2} + \frac{Q^2}{r^2 M_p^2})^{-1} dr^2 + dr^2 d\Omega^2
\eea
where $M_p = G^{-1/2}$ is the Planck mass.  There are two horizons located at
\bea
r_\pm =  \frac{M}{M_p^2} \pm \sqrt{\frac{M^2}{M_p^4} - 4 \frac{Q^2}{M_p^2} }
\eea
The temperature of the black hole is obtained by analytically continuing this solution into Euclidean space and requiring that there are no deficit angles.  The temperature of the black hole is 
\bea
T_\text{BH} = \frac{r_+ - r_-}{4 \pi r_+^2}
\eea
The radiation results in an energy loss for the black hole.  The energy lost per unit area is
\bea
\frac{dE}{dx^2 dt} = - \sum_i \frac{g_i}{4} \int \frac{dp^3}{(2\pi)^3} \frac{p}{e^{(p+\mu_i)/T_\text{BH}} \pm 1} &=& - \sum_\text{fermions} \frac{g_i T_\text{BH}^4}{4}  f(\mu_i) - \sum_\text{bosons} \frac{g_i T_\text{BH}^4}{4} b(\mu_i) \\
f(\mu_i) = - \frac{3}{\pi^2} \text{Li}_4(-e^{-\mu_i/T_\text{BH}}) &\qquad& 
b(\mu_i) =  \frac{3}{\pi^2} \text{Li}_4(e^{-\mu_i/T_\text{BH}})
\eea
where $g_i$ is the number of degrees of freedom excluding (anti)particles and the sum i goes over both particles and anti-particles.  The chemical potential comes from the charge of the black hole and is $\mu_i =  q_i \Phi = q_i Q/r_+$.  We made the simplifying assumption that the particles involved are massless.  To get the energy lost per unit area from the energy lost per unit volume, we multiply by a factor of 1/4.  1/2 is due to specifying that the radiation go outwards rather than inward and another 1/2 comes from an angular average.  If we look at only the photon, this reproduces the well known Stefan-Boltzmann equation.  The total energy lost per unit time is obtained by multiplying by $4 \pi r_+^2$ to give
\bea
\frac{dE}{dt} &\approx& - \sum_\text{fermions}  \frac{g_i M_p^4}{1024 \pi^3 M^2}  f(\mu_i) - \sum_\text{bosons}  \frac{g_i M_p^4}{1024 \pi^3 M^2} b(\mu_i) \equiv - \frac{K M_p^4}{M^2}
\eea
where we have taken the $\mu \ll T$ limit and $K = g_\star/30720 \pi$ with $g_\star = \sum_\text{fermions} g_i (7/8) + \sum_\text{bosons} g_i$.  
We thus find that the mass as a function of time is
\bea
M(t) = \left ( M_o^3 - 3 K M_p^4 t \right )^{1/3}
\eea
so that the evaporation time is
\bea
\label{Eq: evap}
t = \frac{M_o^3}{3 K M_p^4}
\eea

We now calculate how quickly the black hole loses its charge.  Just like how energy is lost, the black hole loses charge at a rate
\bea
\label{Eq: charged BH}
\frac{dQ}{dt} &=& 4 \pi r_+^2 \sum_i \frac{g_i q_i}{4} \int \frac{dp^3}{(2\pi)^3} ( \frac{1}{e^{(p+\mu_i)/T} + 1} - \frac{1}{e^{(p-\mu_i)/T} + 1} ) \\ \nonumber
&=&  - 4 \pi r_+^2 \sum_i \frac{g_i q_i}{4} ( \frac{\mu_i^3}{6 \pi^2}  + \frac{\mu_i T^2}{6} ) \approx - Q \sum_i \frac{g_i q_i^2 M_p^2}{192 \pi M} + \mathcal{O}(Q^2)
\eea
where i now runs over just particles and not anti-particles.  The $\mu_i^3$ term is simply the fact that due to the chemical potential, the ground state has an occupation number where we fill up a fermi sea until $k_F = \mu$.  The $\mu_i T^2$ term is the additional asymmetry generated by the black hole's non-zero temperature.  
Given the complicated dependence of $r_+$, $\mu_i$ and $T$ on the charge Q, we have done an expansion in the small charge regime.  We see that a charged black hole loses its charge exponentially quickly.  This is to be expected as the electromagnetic force is much much stronger than the gravitational force.

\subsection{A black hole with an explicit chemical potential}

We can easily generalize the previous results to a case where an explicit chemical potential is present.  In the cases considered here, the chemical potential will be small.  Thus the rate of energy loss will be the exact same as the previous case of a black hole with a small charge.  The majority of the energy will be taken away from the black hole by the symmetric portion of the Hawking radiation.  The lifetime of the black hole is given by Eq.~\ref{Eq: evap}.

The global charge carried away from the black hole per unit time can be calculated in the exact same manner as before.  The only difference is that because the black hole has no hair and thus no global charge, the global charge carried away from the black hole does not decrease its non-existent global charge.  The black hole temperature is $T_\text{BH} = M_p^2 / 8 \pi M$.  As before, we have 
\bea
\label{Eq: dndt}
\frac{d Q_{B-L}}{d t} = 4 \pi r_\text{BH}^2 \sum_i \frac{q_i g_i}{4} \frac{\mu_i T_\text{BH}^2}{6} = \sum_i q_i g_i \frac{\mu_i}{96 \pi}
\eea
where $q_i$ is the B-L charge of the particle and the sum over i does not include anti-particles.  In all cases considered in this paper, we have $\mu \ll T_\text{BH}$ so we only kept the leading order piece in $\mu$ tossing out the subdominant $\mu^3$ term.  The total charge asymmetry generated from the decay of a black hole of mass $M_o$ is
\bea
\label{Eq: qtotal}
Q_{B-L} &=& \sum_i q_i g_i \frac{\mu_i}{96 \pi} \frac{M_o^3}{3 K M_p^4} \\ \nonumber
&\sim& \sum_i \frac{320 q_i g_i \mu_i M_o^3}{3 g_\star M_p^4}
\eea
Depending on how the black holes are used to generate a baryon number, both Eq.~\ref{Eq: dndt} and Eq.~\ref{Eq: qtotal} will be useful. The difference between Eq.~\ref{Eq: dndt} and Eq.~\ref{Eq: charged BH} is that the chemical potential is now explicit in the Lagrangian rather than induced by a non-zero electric potential.

\section{Baryogenesis from black hole evaporation after inflation}
\label{Sec: after inflation}

In this section we use the coupling shown in Eq.~\ref{Eq: def2} and repeated here for clarity
\bea
\mathcal{L} \supset \int dx^4 \sqrt{-g} \lambda \frac{\partial_\mu \mathcal{R}}{M_p^2} J^\mu_{B-L}
\eea
This coupling is used to generate B-L from the evaporation of black holes after inflation.  The chemical potential is
\bea
\mu_i = \lambda q_i \frac{\mathcal{\dot R} }{M_p^2} = - 9 \lambda q_i (1 + w ) (1 - 3 w) \frac{H^3}{M_p^2} %= - \sqrt{3} \lambda q_i (1+w)(1-3w) \rho^{3/2} \frac{(8 \pi)^{3/2}}{M_p^5}
\eea
so that the evaporation of black holes gives a non-zero B-L abundance.  In particular, any given black hole emits asymmetric radiation at the rate
\bea
\frac{dQ_{B-L}}{dt}  = - \sum_i g_i q_i^2 \frac{3 \lambda (1+w)(1-3w)}{32 \pi} \frac{H^3}{M_p^2} \equiv c(w) \frac{1}{M_p^2 (t-t')^3}
\eea
where $t'$ is some time offset.  When the universe is radiation dominated, $w=1/3$ so to leading order, the chemical potential vanishes.  At loop level the running of the gauge coupling constants provide a non-zero result\cite{Kajantie:2002wa}
\bea
1-3w = \frac{5}{6 \pi^2} \frac{g^4}{(4 \pi)^2} \frac{(N_c + \frac{5}{4} N_f) (\frac{11}{3} N_c - \frac{2}{3} N_f ) }{2 + \frac{7}{2} (N_c N_f/(N_c^2 -1 ))}
\eea
For the SM, this is around $10^{-3}$.  Throughout the paper, when plotting results, the value of $H_\text{inf} = 10^{14}$ GeV, as suggested by BICEP2\cite{Ade:2014xna}, will be used.  The analytical form will always be given so that the results can be easily extrapolated to smaller values of H.

We will be agnostic about the origin of the black holes used.    They may be formed because the inflaton decayed into black holes or large density perturbations after inflation caused a large number of black holes to be formed out of the ambient radiation.  Depending on the details of their formation, certain number densities and mass spectra will be preferred.

We consider two scenarios.  The first is when the energy density of the black holes is a small compared to the energy density in radiation.  The second scenario is when reheating happens via the decay of black holes.  In both cases, we find that if the black holes have masses much larger than the Planck mass, then the coupling $\lambda$ must be very large.
While it may seem unnatural to have such large values of $\lambda$, having $\lambda \sim 10^{10}$ can easily happen if Eq.~\ref{Eq: def2} is suppressed by the Hubble scale rather than the Planck scale.
Another reason why large $\lambda$ may not be objectionable is that if $B-L$ is conserved perturbatively, then no perturbative physics is sensitive to Eq.~\ref{Eq: def2}.  As only non-perturbative effects are sensitive to $\lambda$, an exponentially large coupling may (however unlikely) be completely natural in the full theory of quantum gravity.

\subsection{Black Holes with small energy density}

In this subsection, we assume that the black holes evaporate quickly and contribute negligibly to reheating.  We assume that after inflation ends, there is a number density of black holes $n_\text{BH} = 3 \epsilon/4 \pi r_\text{BH}^3$.  In this way $\epsilon$ corresponds to the fraction of the universe in a black hole.  $\epsilon = 1$ is a universe that is one big black hole.
It is easiest to express everything in terms of the yield.  As entropy is only generated after reheating and is essentially zero before then, we define the entropy at the end of inflation by rescaling the entropy at reheating
\bea
s(t_0) = s(t_\text{RH}) \frac{a(t_\text{RH})^3}{a(t_0)^3}
\eea
where $t_\text{RH}$ is the reheat time and $t_0 = 2/3 H_\text{inf}$ is the time when inflation ends.  We have an effective yield at the end of inflation
\bea
Y^\text{BH}_0 = \frac{n_\text{BH}}{s}(t_0) = \frac{3}{16} \epsilon \frac{M_p^4 T_\text{RH}}{M^3 H^2_\text{inf}}
%\frac{3}{2} \epsilon \frac{T_\text{RH}}{r_\text{BH}^3 H_{inf}^2 M_p^2}
\eea
where $H_\text{inf}$ is the Hubble constant during inflation and $T_\text{RH}$ is the reheat temperature.

The differential equation governing the production of the baryon number asymmetry is 
\bea
\label{Eq: diffeq}
\frac{ d Y^{B-L} }{dt} = \frac{dQ_{B-L}}{dt} Y^\text{BH} = \frac{dQ_{B-L}}{dt} Y^\text{BH}_0 e^{-\Gamma(t-t_0)}
\eea
At a time $t_\text{RH}$ the universe switches from being matter dominated by the inflaton to radiation dominated by the SM.  
\bea
t_\text{RH} = \sqrt{ \frac{5}{\pi^3 g_\star}} \frac{M_p}{T_\text{RH}^2}
\eea

\begin{figure}[t]
     \centering
     \includegraphics[width=0.5\linewidth]{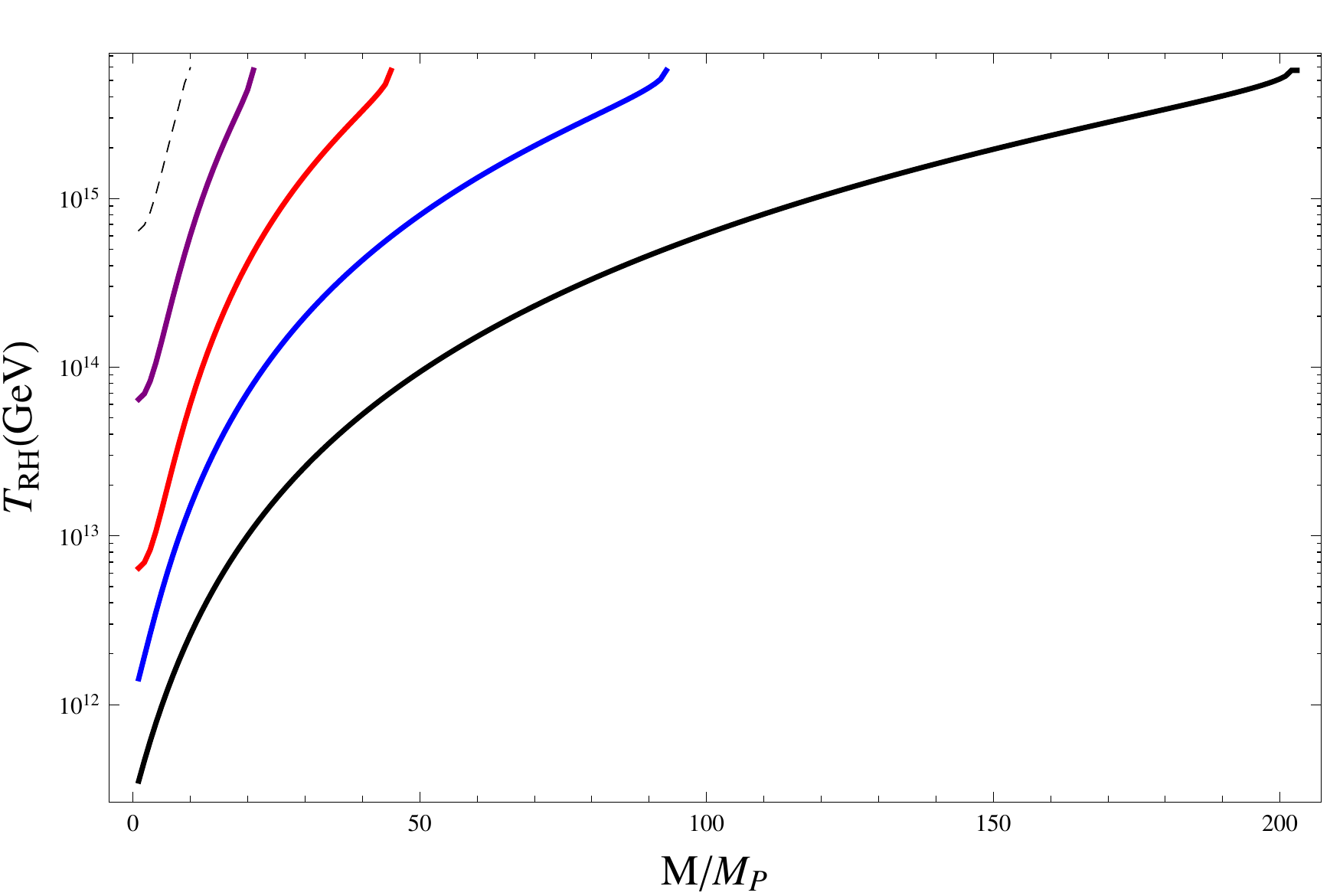}
     \caption{We show the reheat temperature needed to obtain $Y_{B-L} = 10^{-10}$ as a function of the black hole mass.  The black, blue, red, purple and dashed black lines correspond to $\lambda \epsilon = 100, 10, 1, 0.1$ and 0.01 respectively.  After a certain mass, the reheat temperature goes above the maximal allowed reheat temperature and there is no way to obtain the correct baryon number abundance.}  \label{Fig: black hole reheat}
 \end{figure}

For $t_0 < t < t_\text{RH}$ the exact solution to Eq.~\ref{Eq: diffeq} is
\bea
\label{Eq: matter}
Y^{B-L}_\text{matter}(t) = \frac{c(0) Y_0^\text{BH} }{2 M_p^2 t^2 t_0^2} \left (  e^{-\Gamma (t - t_0)} t_0^2 (t \Gamma -1) + t^2 (   1- t_0 \Gamma + e^{t_0 \Gamma} t_0^2 \Gamma^2 (  \text{Ei}(-t \Gamma) - \text{Ei}(-t_0 \Gamma)   )     ) \right ) 
\eea
where Ei is the exponential integral function.  After entering the radiation dominated period $t > t_\text{RH}$, the solution becomes
\bea
\label{Eq: fullsol}
Y^{B-L}_\text{radiation}(t) = Y^{B-L}_\text{matter}(t_\text{RH}) +  \frac{c(\frac{1}{3}) Y_0^\text{BH} e^{-\Gamma t_\text{RH}} }{18 M_p^2 (4 t - t_\text{RH})^2 t_\text{RH}^2} \big (  4 e^{-\Gamma (t-t_0)} (9 e^{\Gamma t_\text{RH}} t_\text{RH}^2 (4 t \Gamma - 4 - t_\text{RH} \Gamma) \\
\nonumber
+ e^{\Gamma t} (4 t - t_\text{RH})^2 (4-3 t_\text{RH} \Gamma) - 9 e^{\Gamma (t_0 + \frac{3 t_\text{RH}}{4})} t_\text{RH}^2 (4 t - t_\text{RH})^2 \Gamma^2 (\text{Ei}(-\frac{3 t_\text{RH} \Gamma}{4}) - \text{Ei}(- (t - \frac{t_\text{RH}}{4}) \Gamma))   \big )
\eea

The simplest limit to study is when the black holes all decay before reheating.  This is the limit where $\Gamma t_\text{RH} \gtrsim 1$.  Additionally, we require that the small black holes that are introduced have less energy than the energy density of inflation.  These limits correspond to 
\bea
\label{Eq: valid}
\Gamma t_\text{RH} \gtrsim 1 \qquad &\Rightarrow& \qquad M \lesssim 5 \times 10^{19}  \left ( \frac{g_\star}{100} \right )^{1/6} \left ( \frac{10^{16}}{T_\text{RH}} \right )^{2/3} \\
\nonumber
n_\text{BH} M \ll V_\text{inflaton} \qquad &\Rightarrow& \qquad M \gg \frac{\sqrt{\epsilon}}{2} \frac{M_p^2}{H_\text{inf}}
\eea
Applying this limit to Eq.~\ref{Eq: matter} and using the fact that black holes which are more massive than a few Planck masses obey $\Gamma t_0 \lesssim 1$, we find
\bea
\label{Eq: approxBL}
Y^{B-L}(t=\infty) = \frac{c(0) Y_0^\text{BH} }{2 M_p^2 t_0^2} = - \frac{39}{512 \pi} \lambda \epsilon \frac{M_p^2 T_\text{RH}}{M^3}
\eea
where we have used the SM value $\sum_i g_i q_i^2 = 13$.  For black holes more massive than just a few Planck masses, we see that $\lambda \epsilon$ is required to be large in order to obtain the correct baryon number abundance.  Given a value of $M$ and $\lambda \epsilon$, Fig.~\ref{Fig: black hole reheat} uses Eq.~\ref{Eq: fullsol} to calculate the reheat temperature needed to obtain the correct baryon number abundance.  As can be seen, large values of $\lambda \epsilon$ are needed once the mass of the black holes exceeds the Planck mass by a significant amount.
%This conclusion can be avoided if there is additional matter so that $(1-3w)$ in the radiation dominated phase is not small.

\subsection{Reheating from black holes}

In this subsection, we assume that at some point in time most of the energy density of the universe is in black holes.  Assume that we have a number density of black holes $n_\text{BH} = 3 \epsilon/4 \pi r_\text{BH}^3$.  The decays of the black holes are asymmetric due to the coupling shown in Eq.~\ref{Eq: def2}.  Depending on the initial abundance and mass of the black holes, there are two possible situations.  In the first situation, when the black holes are created, the  Hubble constant is larger than the decay rate of the black holes.  Thus before the black holes can decay, their number density is diluted.  Once the Hubble constant falls below the decay rate, the black holes all decay and convert their energy into radiation.  Using this criteria, we find that 
\bea
\label{Eq: BHdecaytime}
\Gamma_\text{BH} = H \quad \Rightarrow \qquad \epsilon = \left (  \frac{g_\star}{5120 \pi} \frac{M_p^2}{M^2}  \right )^2 \approx  \left (\frac{0.08 M_p}{M}  \right )^4
\eea
Thus the density of black holes when they decay is a fixed quantity.
In the other situation, the black holes are created such that $\Gamma_\text{BH} > H$.  From the start, the density of the black holes is smaller than what is shown in Eq.~\ref{Eq: BHdecaytime} so that they decay right away.  Thus we can take Eq.~\ref{Eq: BHdecaytime} as an upper bound on the density of black holes.  In most realistic models of black hole creation, e.g. large density fluctuations, this bound will be saturated.

After all of the energy in the black holes has been converted into radiation, the temperature and entropy of the bath is
\bea
n_\text{BH} M = \frac{\pi^2}{30} g_\star T_\text{RH}^4 \qquad \Rightarrow \qquad T_\text{RH} &=& \left (  \frac{45 \epsilon}{16 \pi^3 g_\star} \frac{M_p^6}{M^2}   \right )^{1/4} \\
s &=& \frac{2 \pi^2}{45} g_\star T_\text{RH}^3
\eea
while the number density of B-L asymmetry generated is
\bea
n_{B-L} = Q_{B-L} n_\text{BH} = - \frac{90}{\pi g_\star} \sum_i q_i^2 g_i \lambda \epsilon H^3 = - \frac{45}{4 \pi g_\star}  \sum_i q_i^2 g_i \lambda \epsilon^{5/2} \frac{M_p^6}{M^3}
%n_{B-L} = Q_{B-L} \frac{3 \epsilon M_p^6}{32 \pi M^3} = - \frac{90}{\pi g_\star} \sum_i q_i^2 g_i \lambda \epsilon H^3 = - \frac{45}{4 \pi g_\star}  \sum_i q_i^2 g_i \lambda \epsilon^{5/2} \frac{M_p^6}{M^3}
\eea
Where $Q_{B-L}$ is the total charge generated per black hole and is given in Eq.~\ref{Eq: qtotal}.  The baryon number abundance is then
\bea
Y_{B-L} = \frac{3^{5/2} 5^{5/4}}{g_\star^{5/4} \pi^{3/4} } \sum_i q_i^2 g_i \lambda \epsilon^{7/4} \left ( \frac{M_p}{M} \right )^{3/2} 
\eea
Using Eq.~\ref{Eq: BHdecaytime}, we see that the B-L asymmetry is
\bea
Y_{B-L} \sim 4 \times 10^{-8} \lambda \left ( \frac{M_p}{M} \right )^{17/2}
\eea
Thus this mechanism only works for very small Planck mass black holes or large couplings $\lambda$.

\section{Baryogenesis from primordial black holes during inflation}
\label{Sec: baryogenesis}

In this section, we show how if black holes are constantly being created during inflation, a B-L asymmetry is naturally generated.   We will be using Eq.~\ref{Eq: def}, repeated below for convenience, as Eq.~\ref{Eq: def2} gives a chemical potential that is many orders of magnitude smaller.
\bea
\mathcal{L} \supset \int dx^4 \sqrt{-g} \frac{\partial_\mu \phi}{M_\star} J^\mu_{B-L}
\eea

  \begin{figure}[t]
     \centering
     \includegraphics[width=0.5\linewidth]{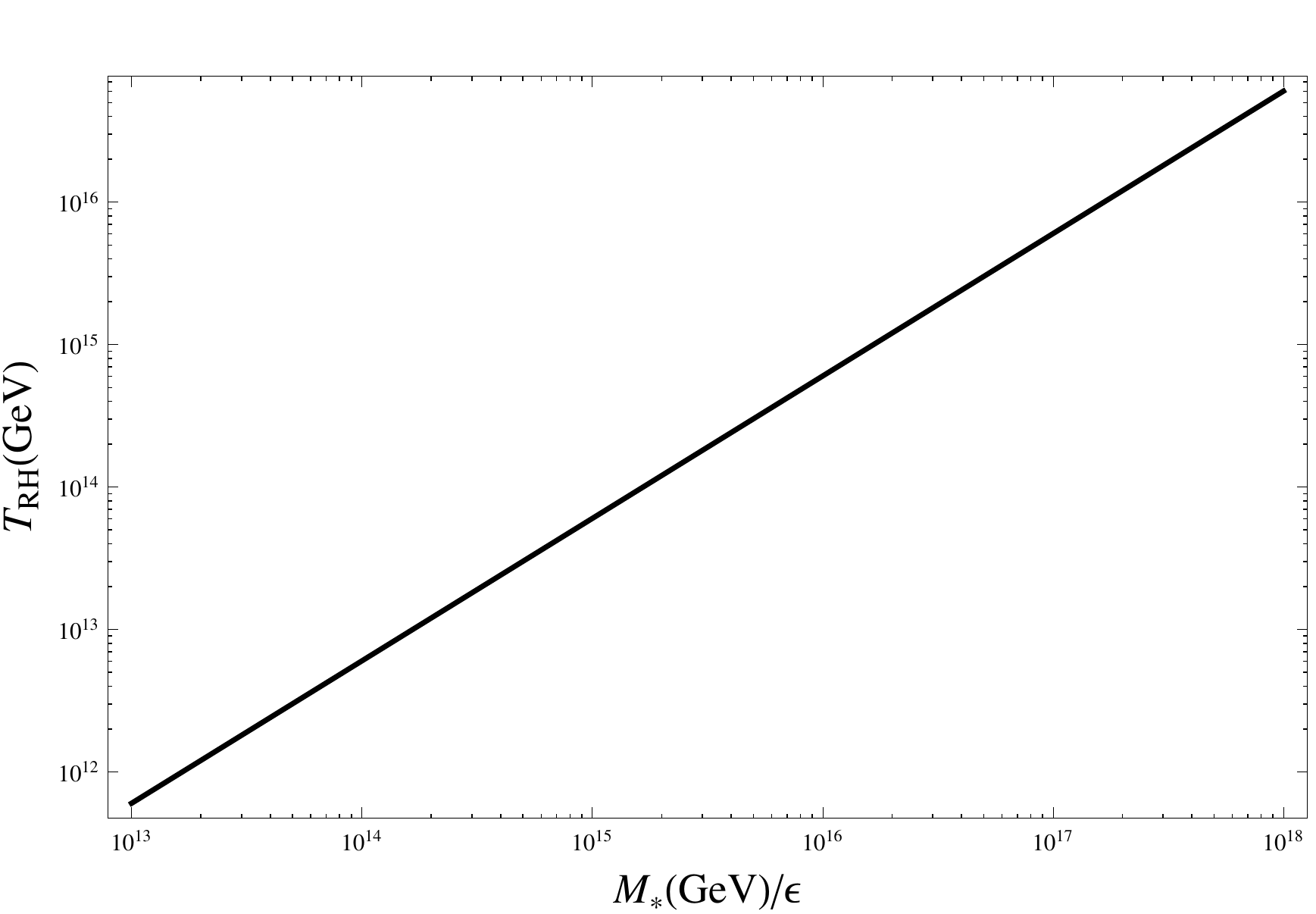}
     \caption{We show the reheat temperature needed to obtain the correct baryon asymmetry as a function of $M_\star/\epsilon$ assuming that the black holes are constantly being produced during inflation.}  \label{Fig: plot}
 \end{figure}

Let us assume that during inflation, small black holes of mass M are being generated at a rate $\frac{3}{4 \pi} \epsilon H^4$.  
The differential equation governing the number density of black holes is
\bea
\frac{dn_\text{BH}}{dt} &=& - (3 H + \Gamma) n_\text{BH} + \frac{3}{4 \pi} \epsilon H^4 \\
n_\text{BH}^\text{eq} &=& \frac{3}{4 \pi}  \frac{\epsilon H^4}{3 H + \Gamma}
\eea
The black holes are constantly evaporating and generating a B-L asymmetry.  The B-L asymmetry being generated is
\bea
\frac{dn_{B-L}}{dt} &=& \frac{d Q_{B-L}}{d t} n_\text{BH} - 3 H n_{B-L} \\
\label{Eq: neq}
n_{B-L}^{eq} &=& \frac{d Q_{B-L}}{d t} \frac{n_\text{BH}^{eq}}{3 H}  =  \sum_i q_i g_i \frac{\mu_i}{96 \pi} \frac{\epsilon H^3}{4\pi (3 H + \Gamma)}
\eea
where we take $\frac{d Q_{B-L}}{d t}$ from Eq.~\ref{Eq: dndt} and i runs over all particles excluding anti-particles.  After inflation ends, a baryon number asymmetry will be observed.  It is mainly produced in the last e-folding of inflation.

If we take the interaction given in Eq.~\ref{Eq: def}, the chemical potential has the value 
\bea
\mu_i = \frac{q _i \dot \phi}{M_\star} = \frac{q_i}{8} \sqrt{\frac{r}{\pi}} \frac{H M_p}{M_\star}
\eea
where $r$ is the scalar to tensor ratio, $q_i$ is the B-L charge of the particle and single field inflation has been assumed\footnote{See \cite{Baumann:2009ds} for a review of inflation and references.}.  The total baryon number density during inflation is given in Eq.~\ref{Eq: neq}.  
After inflation ends, the universe becomes inflaton dominated until reheating has finished.
\bea
\rho_\text{inf} &=& \frac{3 H^2 M_p^2}{8 \pi} (\frac{a_i}{a_f})^3  =  \frac{\pi^2}{30} g_\star(T_\text{RH}) T_\text{RH}^4 \\
n_\text{B-L}(T_\text{RH}) &=& n_\text{B-L}^{eq} (\frac{a_i}{a_f})^3
\eea
The final B-L abundance is
\bea
\label{Eq: yield}
Y_\text{B-L} &=& \frac{n_\text{B-L}(T_\text{RH})}{s(T_\text{RH})} =  \frac{1}{4608 \pi} \left ( \sum_i q_i^2 g_i \right ) \sqrt{\frac{r}{\pi}} \frac{\epsilon}{1 + \frac{\Gamma}{3H}} \frac{H T_{RH}}{M_p M_\star}
\eea
As this equation has many dependencies, we make two simplifying assumptions for the sake of plotting the result.  We assume that $\Gamma \ll H$, that only the SM carries B-L charge ($\sum_i q_i^2 g_i = 13$) and $r=0.16$.  With these assumptions, for any given value of $M_\star/\epsilon$, we can find what the re-heating temperature needs to be in order to have generated the observed baryon number asymmetry.  The reheating temperature as a function of $M_\star/\epsilon$ is shown in Fig.~\ref{Fig: plot}.  Given that we know that $H_\text{inf} \lesssim 10^{14}$ GeV, the maximum reheat temperature is 
\bea
T_\text{RH,max} = (\frac{45 H^2 M_p^2}{4 \pi^3 g_\star(T_\text{RH})})^{1/4} \sim 9 \times 10^{15} \, \text{GeV} 
\eea
and is reached for $M_\star/\epsilon \sim 9 \times 10^{17}$ GeV.
For smaller reheat temperatures, we have $T_\text{RH} \sim 10^{10}$ GeV and $M_\star/\epsilon \sim 10^{12}$ GeV or $T_\text{RH} \sim 10^{12}$ GeV  and $M_\star/\epsilon \sim 10^{14}$ GeV.  
 
\begin{figure}[t]
     \centering
     \includegraphics[width=0.5\linewidth]{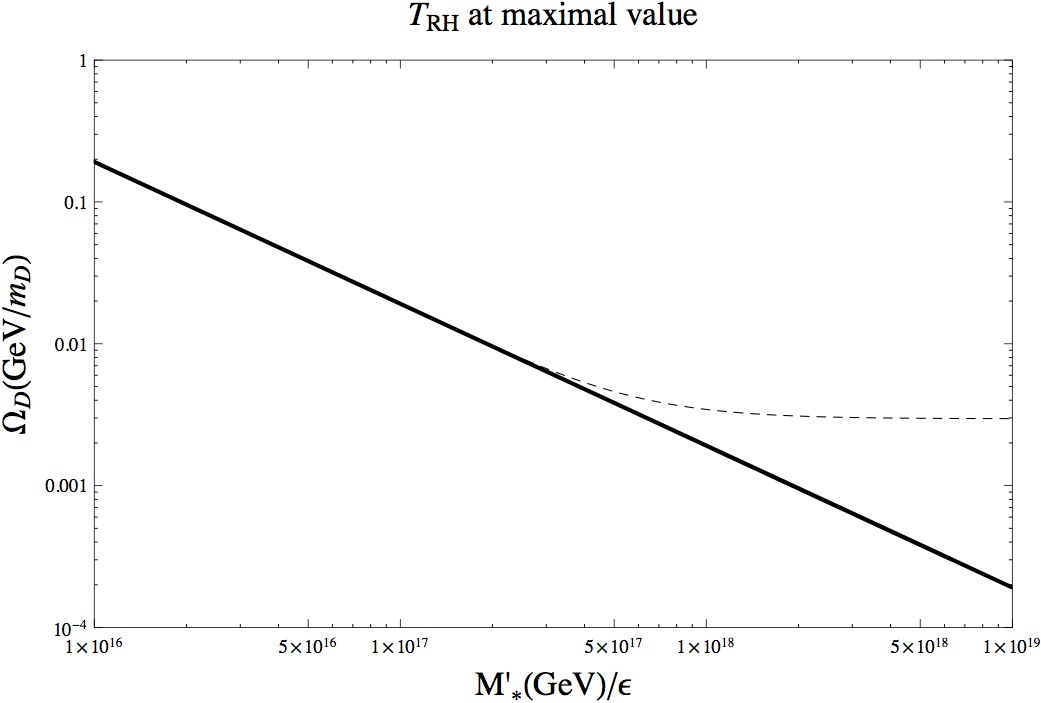}
     \caption{We show the abundance of a 1 GeV dark matter as a function of the suppression scale over the density of black holes ($M'_\star/\epsilon$).  It is assumed that the dark matter is a single fermion with charge 1, that the primordial black holes have a mass of $10^4 M_p$ and that one primordial black hole is created per e-folding.  The thick line shows the difference between particle and anti-particle abundances, while the dashed line shows the sum.  The abundances scales linearly with the mass, reheating temperature and Hubble constant during inflation so that abundances for other masses, reheating temperatures and Hubble constants can be easily extracted.}  \label{Fig: plot2}
 \end{figure}
Dark matter abundance as a function of suppression scale is shown in Fig.~\ref{Fig: plot2}.  If the dark matter can annihilate, then only the asymmetric portion (the thick line) will remain.  The symmetric portion (the dashed line) is more model dependent as it depends on the mass of the black hole, $\lambda'$, $\epsilon$ and the reheat temperature.  In making Fig.~\ref{Fig: plot2} we assumed that $\epsilon = 1$, the reheat temperature is the maximal value and that the mass of the black hole was $10^4 M_p$.

\subsection{Constraints}

In this subsection, we comment on the various constraints/assumptions that the were used.  
We are studying black holes in the first $t \sim 1/H$ seconds of their life.  If the black holes are small, then they reach their equilibrium state in a time $t_\text{eq} \ll 1/H$ so that the thermal approximation is valid.  These black holes evaporate as if they were in flat space and the analysis before goes through unchanged.  

There are two issues that arise when the black holes become larger.  The first is that as their radius approaches Hubble, the time it takes for them to equilibrate is $t_\text{eq} \sim 1/H$.  Thus we are studying black holes out of equilibrium.  We have no good understanding of black holes out of equilibrium.  The previous results may change by only $\mathcal{O}(1)$ or they may change by much more.  We have no way of quantifying the behavior of non-equilibrium large black holes.

Even if the large black holes reach their steady state, they start absorbing the Gibbings Hawking radiation\cite{Gibbons:1977mu} reducing the rate at which an asymmetry is generated.  In de Sitter space, black holes have the Lorentzian Schwarzschild-de Sitter metric
\bea
ds^2 = -\left ( 1 - \frac{2 m}{M_p^2 r} - H^2 r^2 \right ) dt^2 + \left ( 1 - \frac{2 m}{M_p^2 r} - H^2 r^2 \right )^{-1} dr^2  + r^2 d \Omega_2^2
\eea
there are two roots which represent the cosmic and black hole horizons.  At a critical mass $m = M_p^2 / 3 \sqrt{3} H$, the two horizons become degenerate.  For larger masses/radius, black holes do not exist.  As shown in Ref.~\cite{Bousso:1997wi}, near maximally sized black holes can behave in a rather strange manner.  Depending on initial conditions, they can either evaporate and shrink in size or they can anti-evaporate and grow in size.
Thus the previous analysis follows through for small black holes while for larger black holes it is unclear whether the previous results are changed by a small amount or a large amount.

The density of the black holes is limited by one black hole per size of the black hole.  This leads to the bound
\bea
n_\text{BH}^\text{eq} \frac{4 \pi r_\text{BH}^3}{3} \ll 1 \qquad \Rightarrow \qquad \epsilon \ll \frac{M_p^6}{M^3 H^3} %\sim \left ( \frac{10^5 M_p}{M} \right )^3
\eea
If there are many smaller black holes being created during inflation, the black holes should not dominate the energy density of the universe.  In order for that to happen, we need
\bea
\label{Eq: energy density}
n_\text{BH}^\text{eq} M \ll V_\text{inflaton} \qquad \Rightarrow \qquad  \epsilon \ll \frac{M_p^2}{M H}
\eea

Finally, the evaporation of black holes after inflation places constraints.  If the black holes evaporate fast, then they do not inject significant amounts of entropy into the system and the previous results are correct.  In the worst case, there may be large entropy injections that change the result.  Initially, the energy density in the black holes is much smaller than the energy density of the inflaton.  Before reheating, the energy density in the black holes and inflaton both redshift away as $a^{-3}$.
Before the inflaton decays, the universe is matter dominated ($a \sim (H t)^{2/3}$).  At the time of reheating, all of the energy density in the inflaton goes into reheating the standard model and we have
\bea
\frac{V_\text{inflation}}{a(t_\text{RH})^3} \sim T_\text{RH}^4 \qquad \Rightarrow \qquad t_\text{RH} \sim \frac{M_p}{T_\text{RH}^2}
\eea
where $t_\text{RH}$ is the time of reheating.  At the time of reheating, the energy density in the black holes is
\bea
\rho_\text{BH}(t_\text{RH}) = \frac{n_\text{BH}^\text{eq}}{a(t_\text{RH})^3} M \sim \epsilon \frac{ M H}{M_p^2} T_\text{RH}^4
\eea
After reheating, the universe is radiation dominated.  In order for Eq.~\ref{Eq: yield} to be correct, the black holes must evaporate before they become a significant portion of the energy density.  If not then when they evaporate, they inject a significant amount of entropy into the system invalidating the previous results.  The temperature of the SM at the time of evaporation is
\bea
T_\text{evap} \sim T_\text{RH} \left( \frac{t_\text{RH}}{t_\text{evap}} \right )^{1/2} \sim T_\text{RH} \left( \frac{M_p^4}{ M^3 } \right )^{1/2}  \left( \frac{M_p}{ T_\text{RH}^2 } \right )^{1/2} \sim \frac{M_p^{5/2}}{M^{3/2}}
\eea
At the temperature $T_\text{evap}$, the energy density of the black holes is
\bea
\rho_\text{BH}(t_\text{evap}) = \rho_\text{BH}(t_\text{RH}) \frac{a(t_\text{RH})^3}{a(t_\text{evap})^3} \sim \epsilon \frac{ M H}{M_p^2} T_\text{RH} T_\text{evap}^3 
\eea
requiring that this is smaller than the SM energy density gives
\bea
\label{Eq: reheat bound}
\rho_\text{BH}(t_\text{evap}) \ll T_\text{evap}^4  \qquad \Rightarrow \qquad \epsilon \ll \frac{M_p^{9/2}}{M^{5/2} T_\text{RH} H}
\eea
As the final asymmetric abundance (Eq.~\ref{Eq: yield}) is independent of M, there exist regions of parameter space where the mass of the black hole is chosen such that these constraints are satisfied.

\section{Production mechanisms for primordial Black Holes}
\label{Sec: production}

In this section, we briefly mention a few mechanisms which may produce the primordial black holes used for baryogenesis.  
%We first comment on how in models of hybrid inflation, black holes are very naturally produced right after inflation ends due to large density fluctuations.  The second mechanism we mention involves transitions between two de Sitter spaces of near equal Hubble.  These transitions give a natural mechanism for generating black holes during inflation.
The first source of black holes we mention comes from models of hybrid inflation\cite{GarciaBellido:1996qt}.  The end of hybrid inflation is triggered by the waterfall field obtaining a negative mass and rolling quickly to its minimum.  This process can create large density fluctuations $\delta \rho/\rho \sim 1$.  When these large density fluctuations re-enter the horizon, they can take the particles in thermal equilibrium and squeeze them into a black hole.   This mechanism can provide a large number of black holes at/near the end of inflation.

%The natural mass of the black holes produced by hybrid inflation is the thermal energy inside a Hubble volume.  If we assume that reheating happens instantaneously, we have black holes of size
%\bea
%M = \frac{4 \pi}{3 H^3} \frac{ 3 H^2 M_p^2}{8 \pi} \sim 10^5 M_p
%\eea
%In the context of Sec.~\ref{Sec: after inflation}, we find numerically that the correct abundance requires $\lambda \epsilon \sim 10^{10}$.  This means that the coupling shown in Eq.~\ref{Eq: def2} is suppressed by the Hubble/GUT scales rather than the Planck scale.

Small density fluctuations can also evolve into black holes due to gravitational attraction\cite{Carr:1974nx}.  If the density fluctuations produced during inflation are very blue, then primordial black holes can be produced by the gravitational collapse of the over dense regions\cite{Carr:1975qj}.

%If the reheating process is not uniform but instead has some density fluctuations, it is possible that small black holes will form from particles being produced too closely to each other by chance.  Additionally, small density fluctuations can also evolve into black holes due to gravitational attraction.  If the density fluctuations produced during inflation are very blue, then primordial black holes can be produced by the gravitational collapse of the over dense regions\cite{Carr:1975qj}.

The production of black holes during inflation is significantly more difficult than producing them afterwards.  The reason is that during inflation, the expansion of the universe is too fast to allow for gravitational collapse of particles.  Thus the black holes must be produced directly.
Many models of string theory based inflation involve periodic brane collisions, e.g. axion monodromy\cite{Silverstein:2008sg}.  These brane collisions might periodically generate black holes during inflation giving a constant source of black holes throughout inflation.  While warm inflation models have focused on production of particles, it is equally possible that instead of producing particles, a heavy inflaton field may decay into black holes instead.

A final amusing source of black holes is the fact that de Sitter space is actually unstable and decays into black holes.  This decay can be understood as the instability of a finite temperature system to decay into black holes of size $1/T$\cite{Ginsparg:1982rs,Bousso:1996au}.  The decay rate of de Sitter space goes as $\Gamma \sim e^{- 2 \pi M_p^2/3 H^2} \ll 1$.  So we see that this is not a significant source of black holes during inflation.

\section{Conclusion}

In summary, we have shown that black holes evaporate asymmetrically in the presence of a chemical potential.  From an effective field theory perspective, one of the first couplings allowed by symmetries connecting gravity to the SM is a chemical potential.  The expansion of the universe provides T violation while a CP violating interaction with a current allows for the preferential production of baryon number by quantum gravity effects.  

This method of producing the observed B-L asymmetry is unique in that the baryon number violation is provided by quantum gravity effects and not perturbative dynamics.  The utilization of quantum gravity for baryogenesis allows one to do baryogenesis in the context of the Standard Model.  This approach provides an alternative to Electroweak baryogenesis as a way the Standard Model can obtain the observed B-L asymmetry. 
In order for this mechanism to work, there needs to be black holes only a bit heavier than the Planck mass or the couplings involved need to be suppressed by a smaller scale rather than the Planck scale.  

There are several future directions in which to proceed.  The derivation of asymmetric Hawking radiation relied heavily on the analytic continuation of the gravitational solution into Euclidean space.  A more physical picture of Hawking radiation can be provided by using the Bogoliubov transformations between the tortoise and the Kruskal coordinates.  It would be interesting to obtain a derivation of asymmetric Hawking radiation from this perspective.  Another interesting avenue to study would be to better understand the spacial dependence of the chemical potential.  In this paper we assumed that the chemical potential was approximately constant throughout all space and neglected the back reaction of the black hole.  It would be interesting to see when this assumption holds.

%As the BICEP2 measurement of r is already pointing to new physics around the GUT scale, the suppression of these couplings by the GUT scale can be obtained in physically motivated models.

\section*{Acknowledgments}

We thank Daniele Alves, Nima Arkani-Hamed, Spencer Chang, Daniel Harlow, Shamit Kachru, Mariangela Lisanti and Juan Maldacena for stimulating discussions and helpful comments on the draft.  We would like to especially thank Sergei Dubovsky and Neal Weiner for pointing out an important mistake in a previous version of this paper. 
A.H.~is supported by the Department of Energy under contract DE-SC0009988.

\end{document}